\documentclass[epj]{svjour}
\usepackage{amsmath,amsfonts,amssymb,wrapfig,graphicx,footnote}
\usepackage{graphics}
\def\beq{\begin{eqnarray}}
\def\eeq{\end{eqnarray}}

\def\lie{\pounds}
\def\nn{\nonumber\\}

\def\half{{\textstyle{\frac{1}{2}}}}
\begin{document}
\title{Quasilocal first law of black hole dynamics from local Lorentz transformations}
\author{Ayan Chatterjee\inst{1}\thanks{\emph{Email}:ayan.theory@gmail.com}%
\and Avirup Ghosh\inst{2}
\thanks{\emph{Present address:} Indian Institute of Technology, Gandhinagar- 382355, India}%
\thanks{\emph{Email}:avirup.ghosh@iitgn.ac.in}%
}             
%
%
\institute{Department of Physics and Astronomical Science, Central University 
of Himachal Pradesh, Dharamshala -176215, India. \and Theory Division, Saha Institute of Nuclear Physics, 1/AF Bidhan 
Nagar, Kolkata 700064, INDIA.}
\date{Received: date / Revised version: date}
%
\abstract{
Quasilocal formulations of black hole are of immense importance since they reveal the essential and minimal assumptions required for a consistent description of black hole horizon, without relying on the asymptotic boundary conditions on fields.  Using the quasilocal formulation of Isolated Horizons, we construct the Hamiltonian charges corresponding to local Lorentz transformations on a spacetime admitting isolated horizon as an internal boundary. From this construction, it arises quite generally that the \emph{area} of the horizon of an isolated black hole is the Hamiltonian charge for local Lorentz boost on the horizon. Using this argument further, it is shown that, observers at a fixed proper distance $l_{0}$, very close to the horizon, may define a notion of horizon energy given by $E=A/8\pi G l_{0}$, the surface gravity is given by $\kappa=1/l_{0}$, and consequently, the first law can be written in the quasilocal setting as $\delta E=(\kappa/8\pi G)\delta A$.
\PACS{
      {04.70.−s}{Physics of black holes}   \and  
      {04.70.Bw}{Classical black holes}
     } 
} 
\maketitle
\section{Introduction}
\label{intro}
The laws of classical dynamics of black hole horizons have been shown to be identical to the  laws of thermodynamics 
\cite{Bardeen:1973gs,Hawking:1971vc,Hawking:1974sw,Bekenstein:1973ur,Bekenstein:1974ax}. This  
implies that black holes are thermal objects and hence, there must exist a deep connection between 
the dynamics of spacetime (gravity) and thermodynamics of horizons. Hawking's result that black holes can be 
assigned a temperature $T=\kappa/2\pi$, and the first law of black hole mechanics requires that black holes of area 
$A$ must have thermodynamic entropy given by $S=A/4G$ \cite{Hawking:1974sw,Bekenstein:1973ur,Bekenstein:1974ax,Wald:1995yp,Wald:1993nt,Iyer:1994ys}. The search for the statistical mechanical explanation
of this entropy has led to deeper understanding regarding it's origin from the appropriate microscopic states residing on the black hole horizon. Indeed, the microscopic state counting exercise in string theory and in loop quantum gravity,  not only gives the semi- classical Bekenstein- Hawking area law but also gives the quantum mechanically important logarithmic corrections. The detail matching of the miroscopic computation of entropy with the Bekenstein- Hawking area law, is believed to indicate that these two are candidate theories for quantum gravity. However, in absence 
of any fully established theory of quantum gravity, one may try for a deeper understanding of the classical laws of black hole mechanics in search of clues relating thermodynamical quantities with gravitational or geometric quantities. 

It has been argued in \cite{Frodden:2011eb,Bianchi:2012vp}, in the context of stationary black holes, that it may be possible for observers, fixed at a proper distance $l_{0}$ from the horizon, to define a notion of \emph{energy} ($E$) proportional to the \emph{horizon area} ($A$), given by $E=A/8\pi G l_{0}$. This connection between $E$ and $A$ has been derived using the physical process version of the first law of black hole mechanics where one considers the change in the black hole parameters due to absorption of infalling matter. 
Surprisingly, the relation between energy and 
the horizon area, does not get any contribution from matter fields \cite{Frodden:2011eb}. This is an important result, since it establishes a correlation between geometry and thermodynamics on the horizon, as perceived by a local observer.


Thus, it is worthwhile to check if this relation can also be obtained from the phase space or Hamiltonian perspective. The absence of matter contributions in the above mentioned result hints at the existence of transformations which act non- trivially on the gravitational fields but act trivially on matter fields like the Maxwell fields.  Clearly, these are nothing but the local Lorentz transformations. \footnote{Note that these transformations may however, have a non- trivial action on the fields transforming under half- integral spin representations of the
 Lorentz Lie algebra, but, we shall not consider them here. } 

In order to account for the action of the local Lorentz transformations on the phase- space of general relativity, we must use the first order tetrad- spin-connection formalism since the second order metric formalism is insensitive to them.
Having identified these transformations one can look for a subset of these that keep the black hole boundary conditions intact. Since black hole horizons are non- expanding marginally trapped null surfaces
(NEH), it is therefore natural to look for the local Lorentz transformations which map the NEH to itself. 
As is well known, these must belong to $ISO(2)\ltimes \mathbb{R}$, the 
little group of the Lorentz group \cite{Chatterjee_ghosh_basu} that keeps a null vector invariant. Further, using the covariant phase- space formalism for first order gravity one can derive the expression for
the charges corresponding to these transformations. Since these are local gauge transformations, their bulk Hamiltonian generators are zero by the equations of motion. However, due to presence of a horizon we get non trivial contributions from the boundary. 
In other words, a subgroup of these transformations turn out to be genuine physical symmetries rather than being pure gauge. 
Indeed, we show that a certain class of null Lorentz boosts have non trivial charges and are therefore genuine symmetries. These charges are precisely the area of the horizon. We appropriately adopt the expression for these charges to derive the quasilocal first law discussed in \cite{Frodden:2011eb}.



In the following, we give a description of our arguments used in the subsequent sections and our results.
In section 2, we briefly review the geometry of weak isolated horizon (WIH) and argue that both Rindler and black hole horizons satisfy the boundary conditions of a WIH. This is not surprising since given a spacetime with a non- extremal horizon, all stationary observers at a 
\emph{small distance }$l_{0}$ from the  horizon perceive the near horizon geometry to be Rindler spacetime, the  horizon being generated by the boost Killing vector field of that metric (see \cite{Jacobson:1995ab,Padmanabhan:2013nxa} and references therein).  We use this fact to extend the phase- space of isolated horizons (which already contains black hole spacetimes) to include the Rindler horizons as well. Thus black hole horizons and Rindler horizons belong to the same space of solutions of Einstein's theory which admit a WIH as an inner boundary (this is possibly true for other theories of gravity too, but we do not concern ourselves with those theories here).

In section 3, we obtain the horizon area as a charge corresponding to the \emph{local Lorentz boosts}. 
Note that for Rindler horizons as well as for black hole horizons in general relativity, the area is canonically conjugate to 
the boost Killing vector generating the horizon\cite{Carlip:1993sa,Massar:1999wg,Wall:2010cj,Wall:2011hj}. However, in our case, the horizon area is a charge arising due to the local `Lorentz boosts' and not due to `boost vector fields'. 

In section 4, we obtain the quasilocal first law for black hole mechanics. We use an improved notion of symmetries (called Lorentz- Lie derivative) where one insists that the coordinate invariance be extended to 
allow for non- invariance that can be compensated by the gauge transformation of that gauge field \cite{Jackiw:1979ub,Forgacs:1979zs,Ortin:2002qb,Obukhov:2006ge,Fatibene,Gralla:2014yja}. The action of diffeomorphism
generated by a Killing vector field $\xi^a$ on the tetrad $e^{I}_{a}$ is given by $\lie_{\xi}e^{I}=\epsilon_{(\xi)}{}^{I}{}_{J}\, e^{J}$, 
with $\epsilon_{(\xi)}{}_{IJ}$ being a local Lorentz transformation matrix.
While the tetrads adapted to the horizon are Lie dragged by the boost Killing vector, the tetrads adapted to the stationary observer, just outside the horizon, are not. This is because  they are boosted with respect to those adapted to the horizon. The Lorentz- Lie derivative can then be effectively used to find this relative boost.
Likewise, on the isolated horizon phase- space, a local 
boost Lorentz transformation gets fixed through the action of the horizon generating spacetime boost vector field. Besides, since the 
horizon generating boost vector field is a time translational on the horizon, the corresponding
Hamiltonian should rightly be regarded as energy. This leads the relation between the horizon area and energy and to the quasilocal first law for black hole mechanics.

Finally, we would like to point out that the approach that we develop below, to construct the Lorentz charges and hence the quasilocal first law of black hole mechanics, to our knowledge, has not been dealt with before in the literature. There are constructions of Lorentz charges in the context of asymptotically locally AdS spaces in \cite{Aros:1999id,Aros:1999kt,Aros:2002ub} and horizon Noether charge for a combination of diffeomorphism and  local Lorentz transformations in \cite{Jacobson:2015uqa} in the context of  black hole entropy. Further, such constructions have also been considered in the context of black hole mechanics \cite{prabhu,Setare:2015nla,Setare} However, our approach differs substantially both in motive and in methodology.

\section{Isolated horizon as the inner boundary of spacetime}
The formalism of isolated horizons is useful to model a black hole horizon. They are black hole analogues
of isolated equilibrium states in thermodynamics and correspond to black hole horizons which are not interacting with the surroundings.
An Isolated horizon does not require the existence of time like Killing vector fields in the neighbourhood of the horizon, only the intrinsic and 
extrinsic geometry is enough. It is defined to be a null hypersurface which is expansion free, having constant surface gravity and on which the Einstein field equations hold. 
It has found important application in establishing the quasilocal \footnote{The word quasilocal refers to the fact that the first law and the charges which appear in the first law for the horizon requires not only the cross-section of the horizon but also a finite element of the horizon, that is not only a point but also a finite neighbourhood.} definition of mass ($M$)
and angular momentum ($J$) of black holes, and in the proof of the first law of black hole mechanics \cite{Ashtekar:1998sp,Ashtekar:2000sz,Ashtekar:2000hw,Chatterjee:2008if}. 

We first describe the geometrical 
set- up and the boundary conditions. Let us
consider a $4$-manifold $\cal M$ equipped with a metric $g_{ab}$ having signature $(-,+,+,+)$.
Let ${\Delta}$ be a null hypersurface in $\cal M$ generated by a future directed null vector field $\ell^a$. The coordinate system on (a coordinate patch of) $\Delta$ is given as follows. Consider a cross- section $S_{0}$ of $\Delta$ with coordinates $x_{\perp}^{i}\, \, (i=1,2)$ . The tangent on this patch is then given by 
$\ell^{a}=(\partial/\partial \lambda)^{a}$, with $\lambda$ being the affine parameter. Let us, without any loss of
generality, choose the value of the affine parameter on $S_{0}$ as $\lambda=0$. Further, we denote
the spatial cross- sections which foliate the horizon by $S_{\lambda}$ which are essentially surfaces of constant $\lambda$. Thus, if $P$ is any point on $S_{\lambda}$, it's coordinates are $(\lambda, x_{\perp}^{i})$, where 
$\lambda$ is the affine separation of the point $P$ from $S_{0}$.
We are however interested in a specific class of isolated horizons or null surfaces generated by 
a Killing vector field $l^{a}=(\partial/\partial v)^{a}$, existing only on the horizon. The generators $l^{a}$ are null and hence
the parallel transport of $l^{a}$ is also proportional to $l^{a}$. Then, $l^{a}\nabla_{a}l^{b}=\kappa_{(l)}l^{b}$,
where $\kappa_{(l)}$ is the acceleration corresponding to the null normal $l^{a}$. In the context of black hole, this plays the role of surface gravity. If surface gravity is constant, the horizon generating parameter $v$ is
related to the affine parameter $\lambda$ through $\lambda=a\,e^{\kappa\, v}+b$.
Note that if $\xi$ is a positive function on the null surface, then $\xi l^{a}$ is also horizon generating. 
The expansion $\theta_{(l\,)}$ of the null normal $l^a$ is defined by
$\theta_{(l\,)}=q^{ab}\nabla_a l_b$, where $\nabla_a$ is the covariant
derivative compatible with $g_{ab}$ and $q^{ab}$ is the metric on $S_{v}$.

The surface $\Delta$, equipped with the class $[\xi l^a]$ of null normals, is called 
a \textit{weak isolated horizon} (WIH) in $({\cal M},g_{ab})$ if the following conditions hold 
\cite{Ashtekar:1998sp,Ashtekar:2000sz,Ashtekar:2000hw,Chatterjee:2008if}:
\begin{enumerate}\label{bcond}
\item $\Delta$ is topologically $S^2 \times \mathbb{R}$.
\item The expansion $\theta_{(l)}=0$.
\item The equations of motion hold on $\Delta$ and the vector field $-T^a{}_b l^b$
is future directed and causal on $\Delta$.
\item There exists an one form $\omega^{(l)}$ on $\Delta$ which is lie dragged $\lie_{l}\omega^{(l)}=0$.
\end{enumerate}
All these boundary conditions are intrinsic to $\Delta$ and also
imply the existence of a Killing vector field $\xi l^{a}$ on $\Delta$. The first condition
is only a topological restriction while the second condition, that the horizons be expansion free, applies to
black hole horizons and also to Rindler horizons.  The third condition ensures that
equations of motion and energy condition hold. Note that the first three conditions hold true for all vectors in the 
equivalence class $[\xi l^{a}]$. It may be easily inferred from the above conditions that the one- form
$\omega^{(l)}_{a}=\!-\kappa_{(l)}\,n_{a}+\bar{\pi} m_{a}+\pi \bar{m}_{a}$ exists on $\Delta$. The fourth condition indicates that the one 
form is Lie dragged on the horizon. This condition also ensures that the surface gravity $\kappa_{(l)}$ is a constant on the horizon.
We use the null tetrad $(l, n, m, \bar{m})$ such that $1\!=\!-n\cdot l=\!m\cdot\bar m$ and all other scalar
products vanish. This basis is especially suited for the setup since one of the
null normals $l^a$ matches with one of the basis vector. In this basis the spacetime metric is 
given by $g_{ab}=-2l_{(a}n_{b)}+ 2 m_{(a} \bar m_{b)}$.

These conditions hold not only for black hole hole horizons but also for Rindler horizons. For a stationary observer at a small distance $l_0$ from a non-extremal black hole horizon perceives it as a Rindler horizon. This Rindler horizon is also expansion 
free and furthermore the surface gravity is also constant and hence also satisfy the boundary conditions 
of a WIH. This is the key similarity between a Rindler horizon and a black hole horizon. \footnote{ Note that in a Minkowski spacetime, the Rindler
horizons for accelerated observers also acts as information barriers (in the classical sense). However, obstruction of information is observer dependent; the Rindler horizon 
does not block information from an inertial observer just as a freely falling observer can access the inside of a black hole by crossing it.}

Now, given that the internal boundary is an isolated horizon, one may construct the space of solutions which admit 
an isolated horizon as an internal boundary. For this, we work with the first order Palatini Lagrangian and the covariant phase space formalism. Given a Lagrangian, the on-shell variation gives $\delta  L=d\Theta(\delta)$ 
where $\Theta$ is called the symplectic potential. It is a $3$-form in space-time and a 
$1$-form in phase space. Using this symplectic potential,  one constructs the symplectic current $J(\delta_1,\delta_2)= \delta_1\Theta(\delta_2)-\delta_2\Theta(\delta_1)-\Theta([\delta_1,\delta_2])$, 
which, by definition, is closed on-shell. The symplectic structure is then 
defined to be:
\begin{equation}
\Omega(\delta_1,\delta_2)=\int_{M}J(\delta_1,\delta_2)
\end{equation}
where $M$ is a space-like hypersurface. Since $dJ=0$ provided the equations of motion and linearized equations of motion hold, this implies that when integrated over a closed region of spacetime bounded by $M_+\cup M_-\cup 
\Delta$ (where $\Delta$  is the inner boundary considered),
\begin{equation}
\int_{M_+}J-\int_{M_-}J~+~\int_{\Delta}J=0,
\end{equation}
where $M_+,M_-$ are the initial and the final space-like slices, respectively. For the case when WIH is an internal boundary, third term is exact, $\int_{\Delta}J=\int_{\Delta}dj $, and the hypersurface independent symplectic structure is given by:
\begin{equation}
\Omega(\delta_{1}, \,\delta_{2})= \int_MJ-\int_{S_\Delta}j
\end{equation}
where $S_\Delta$ is the 2-surface at the intersection of the hypersurface  $M$ with 
the boundary $\Delta$. 
The quantity $j(\delta_1,\delta_2)$ is called the boundary symplectic current 
and it's integral, the boundary symplectic structure.

As we have said before, we are interested in constructing the space of solutions of general relativity, 
and we shall use the first order formalism in terms of tetrads and connections. 
This formalism is naturally adapted to the nature of the problem in the sense 
that the boundary conditions are easier to implement, construction of the covariant phase- space becomes simpler
and the action of local Lorentz transformations are easier to disentangle.
For the first order theory, we take the fields on the manifold to be
($e_{a}{}^{I},\, A_{aI}{}^{J}$), where $e_{a}{}^{I}$ is the co- tetrad, $A_{aI}{}^{J}$
is the gravitational connection. The Palatini action in first order gravity is given by \cite{Ashtekar:2000hw,Chatterjee:2008if}:
\begin{equation}\label{lagrangian1}
S_{G}=-\frac{1}{16\pi G}\int_{\mathcal{M}}\left(\Sigma^{I\!J}\wedge 
F_{I\!J}\right)
\end{equation}
where $\Sigma^{IJ}=\half\,\epsilon^{IJ}{}_{KL}e^K\wedge e^L$, $A_{IJ}$ is a Lorentz $SO(3,1)$ connection 
and $F_{IJ}$ is a curvature two-form corresponding to the connection given by
$F_{IJ}=dA_{IJ}+A_{IK}\wedge A^{K}~_{J}$. Our strategy shall be to construct the symplectic structure for the action given in eqn. \eqref{lagrangian1}. Let us first look at the Lagrangian for gravity. The symplectic potential in this case is given by, $16\pi G\,\Theta(\delta)=-\Sigma^{I\!J}\wedge \delta A_{I\!J}$. The symplectic structure is given by \cite{Ashtekar:2000hw,Chatterjee:2008if},
\begin{equation}\label{symplectic_current1}
\Omega(\delta_1,\delta_2)=-\frac{1}{8\pi 
G}\,\int_{M}\delta_{[1}\Sigma^{IJ}\!\wedge\delta_{2]}A_{IJ}
 -\frac{1}{4\pi G}\int_{S_{\Delta}}\delta_{[1}\psi\, \delta_{2]}{}^2\epsilon.
\end{equation}
The function $\psi$ is a potential for the surface gravity $\kappa_{(l)}$ and is defined by $\lie_l\psi=\kappa_{l}$
and ${}^2\epsilon$ is the area two form on the spherical cross sections $S_{\Delta}$ of the horizon. The field 
$\psi$ is assumed to satisfy the boundary condition that $\psi=0$ at some initial cross section.

%
%

%
\section{Lorentz transformations and charges on $\Delta$}
We shall work in the first order tetrad- connection formulation of gravity. This is essential since the second order
metric variables are not ideally suited to study the transformations associated with local Lorentz transformations. In the previous section, we have used this first order 
theory to construct the symplectic structure. To evaluate the charges arising due to 
local Lorentz transformations,  we take a local basis consisting of the co-tetrads $e^I$. The co-tetrads and the connection transform under a Lorentz transformation in the following way.
\beq
e^{I}&\rightarrow &\Lambda^{I}{}_{J}\,e^{J}\\
A^{IJ}&\rightarrow& (\Lambda^{-1})^{I}{}_{K}\,A^{KL}\,\Lambda_{L}{}^{J}+ (\Lambda^{-1})^{I}{}_{K}\,d\Lambda^{KJ}
\eeq
where $\Lambda^I~_{J}$ is the Lorentz transformation matrix. The variations of 
the co tetrads and the connection due to infinitesimal Lorentz transformations, $\Lambda^{I}{}_{J}=(\delta^{I}{}_{J}+\varepsilon \,\epsilon^{I}{}_{J})$, 
are given by (note that $\epsilon ^I_J$ are the generators of the Lorentz transformations and $\epsilon$ is a book keeping parameter for 
the order of variation),
\beq
\delta_\epsilon e^I&=&\epsilon^{I}{}_{J}\,e^{J}\\
\delta_\epsilon A^{IJ}&=&d\epsilon^{IJ} +A^{IK}\epsilon_{K}^{J}+A^{JK}\,\epsilon^{I}_{K}.
\eeq
We also require the expression for the variation of $\Sigma_{IJ}$. After a bit of algebra, one can show that,
\beq
\delta_\epsilon \Sigma_{IJ}
&=&\varepsilon_{IJKL}\,\epsilon^{K}{}_{M}\,e^{M}\wedge e^{L}\nn
&=&\epsilon^{K}{}_{J}\,\Sigma_{I}{}^{K}-\epsilon^{K}{}_{I}\,\Sigma_{J}{}^{K}.
\eeq
%
The action of the Lorentz transformations on the fields in the bulk symplectic structure is obtained as follows:
\beq
\Omega_{B}(\delta_\epsilon,\delta)&=-&\frac{1}{16\pi G}\int_{\mathcal{M}}(\epsilon^K~_J\Sigma_{IK}-\epsilon^K~_I\Sigma_{JK})\wedge\delta A^{IJ}\nn
&-&\delta\Sigma_{IJ}\wedge(d\epsilon^{IJ} +A^{IK}\epsilon_{K}{}^{J}+A^{JK}\,\epsilon^{I}{}_{K})\nn
\eeq
The third term may be rewritten as 
\beq
&\delta\Sigma_{IJ}\wedge d\epsilon^{IJ}=d(\delta\Sigma_{IJ}\,\epsilon^{IJ})-\delta \left(d\Sigma_{IJ}\right)\epsilon^{IJ}\nn
&=d(\delta\Sigma_{IJ}\,\epsilon^{IJ})+\delta(A_I{}^K\wedge\Sigma_{KJ}+A_J{}^K\wedge\Sigma_{IK})\epsilon^{IJ}
\eeq
Therefore, the contribution only survives on the cross- sections of the horizon
\beq\label{bulk_term_charge}
\Omega_{B}(\delta_\epsilon,\delta)&=&\frac{1}{16\pi G}\int_{S_{\Delta}}\delta\Sigma_{IJ}\,\epsilon^{IJ}.
\eeq

To evaluate the above expression, we must determine the form of $\epsilon_{IJ}$. Note that the WIH reduces the 
local Lorentz group $SL(2,C)$ to $ISO(2)\ltimes \mathbb{R}$, the little group of the Lorentz group \cite{Chatterjee_ghosh_basu}. More precisely, the WIH boundary conditions are invariant under a subgroup of the local Lorentz group.  
Explicitly, the Lorentz matrices associated with the transformations which keep the WIH boundary conditions invariant are given by
\begin{align} \Lambda_{IJ}=&-\xi l_In_J-\xi^{-1}n_Il_J+2m_{(I}\bar
m_{J)},\label{L1}\\
\Lambda_{IJ}=&-2l_{(I}n_{J)}+(e^{i\theta}m_I\bar m_J+c.c.),\label{L2}\\
\Lambda_{IJ}=&-l_In_J-(n_I-cm_I-\bar cm_I+|c|^2l_I)l_J\nonumber\\
&+(m_I-\bar cl_I)\bar m_J+(\bar m_I-cl_I)m_J\label{L3}\end{align}
The generators coresponding to these transformations are obtained to be:
\begin{align} &B_{IJ}=(\partial\Lambda_{IJ}/\partial\xi)_{\xi=1}=-2l_{[I}n_{J]},
\label{lbb}\\
&R_{IJ}=(\partial\Lambda_{IJ}/\partial\theta)_{\theta=0}=2im_{[I}\bar
m_{J]},\label{lbr}\\
&P_{IJ}=(\partial\Lambda_{IJ}/\partial{\rm Re}\,c)_{c=0}=2m_{[I}l_{J]}+2\bar
m_{[I}l_{J]},\label{lbp}\\
&Q_{IJ}=(\partial\Lambda_{IJ}/\partial{\rm Im}\,c)_{c=0}=2im_{[I}l_{J]}-2i\bar
m_{[I}l_{J]},\label{lbq}\end{align}
where $B$ generates boost on $\Delta$, $R$ generates rotation on the spherical cross- sections of 
$\Delta$. The generators $P,Q$ generate transformations which keep the direction of $l$ and $n$ invariant 
respectively. The $\epsilon_{IJ}$ in equation \eqref{bulk_term_charge} can thus be either of these above four 
generators.
We recall that the expression of $\Sigma_{IJ}$ in terms of a null basis adapted to $\Delta$ is given in 
the following form (when pulled back on the horizon):
\beq
\Sigma^{IJ}=2l^{[I}n^{J]}~^{2}\epsilon+2n\wedge(im~l^{[I}\bar{
m}^{J]}-i\bar{m}~l^{[I}m^{J]})
\eeq
Given the generators of the Lorentz transformations in eqn. \eqref{lbb}-\eqref{lbq}, it immediately follows that 
the bulk term survives only for null boosts $-2 \eta l_{[I}n_{J]}$  with $\eta$ as a finite parameter for the null boost and the Hamiltonian charge $\delta Q$ is given by
\beq\label{Ham_charge}
\Omega_{B}(\delta_{\eta}, \delta)=\delta Q=\frac{\eta}{8\pi G}\delta A
\eeq
where $\eta$ is the parameter of transformation and $A$ is the area of the cross-section. 
If the phase- space is restricted such that the boost parameter $\eta$ is a 
phase- space constant, then the Hamiltonian is obtained to be:
\beq\label{boost charge}
\delta Q=\delta(\frac{\eta A}{8\pi G})
\eeq
We will further assume that $\eta$ is constant on the horizon. This is necessary so that the null normal $l$ transforms within the class $[\xi l]$, discussed before. This is also necessary to keep the boundary condition of a (WIH) intact. Transformation with arbitrary $\eta$ keeps only the boundary conditions upto condition (3) invariant
We shall show in the next section that the natural value of $\eta$ is $1/l_{0}$ and then, the Hamiltonian charge turns out to be $A/8\pi G l_{0}$. This result holds true for an observer who is residing at a fixed proper distance $l_{0}$ away from the horizon. The relationship between
the Hamiltonian and horizon area that appears is more than a coincidence. It shows that the boosts 
which reside on the horizon are the ones which create a physical charge and is related to the area. As we shall show, this fact is related to the fact that for this \textit{near the horizon} observer the internal boosts are related to the spacetime boosts.

To complete the argument, we show below that the boundary symplectic structure does not contribute 
to the Hamiltonian. Let $\psi$ be a potential for $\kappa_{(l)}$ defined by $\lie_l\psi=\kappa_{(l)}$ and hence,
$d\psi =-\kappa_{(l)}n+\alpha_{(l)}m+\bar{\alpha}_{(l)}\bar{m}$. Then,
$d\delta_{\eta}\psi$ can at most depend on $m$ and $\bar{m}$, and can be written as $d\delta_{\eta}\psi =\delta_{\eta}\alpha_{(l)}m+\delta_{\eta}\bar{\alpha}_{(l)}\bar{m}$. Thus, $\lie_{\ell}\delta_{\eta}\psi=0$.
Therefore $\delta_\eta\psi$ is a function only of coordinates on $S^2$ and can be set equal to zero at the 
initial cross-section. On the other hand $\delta_\eta~^2\epsilon=0$. Therefore the 
boundary contribution vanishes i.e $j(\delta_\eta,\delta)=0$.
We also look at contributions from spacelike rotations given by, $l\mapsto l,n\mapsto n,m\mapsto e^{i\theta}m,\label{lor2}$. Under these transformations $\delta\kappa=\delta n=0$ which implies that one can set 
$\delta_\theta\psi=0$  and hence $\delta_\eta~^2\epsilon=0$. Therefore $j(\delta_\theta,\delta)=0$.
Finally, one may also look at null rotations keeping $l$ fixed, given by 
$l\mapsto l,n\mapsto n-cm-\bar c\bar m+c\bar cl,m\mapsto m-\bar cl,\label{lor3}$, for which it follows that
 $\delta_c~^2\epsilon=0$ and $j(\delta_c,\delta)=0$.


\section{Local Lorentz boosts on the horizon and the first law}
In this section, we shall show that the natural value of $\eta$ is $\/l_{0}$ and in the process, try to decipher the 
meaning of the Hamiltonian charge obtained in the previous section in eqns. \eqref{Ham_charge} and \eqref{boost charge}. We 
show below that one needs a better notion of derivatives, called the Lie- Lorentz derivatives, to handle 
fields which are Lorentz- Lie algebra valued spacetime fields. To exemplify, let us consider the electromagnetic one form $A$ and find how it behaves under a diffeomorphism under a vector field $\chi^{a}$. The result is:
\begin{equation}
\lie_{\chi}A=\chi\cdot F +d(\chi\cdot A).
\end{equation}
This Lie derivative is not gauge invariant but can be made so by subtracting the gauge transformation term 
$d(\chi\cdot A)$, if we define a new Lie derivative by
\begin{equation}
\lie^{'}_{\chi}A=\lie_{\chi}A - d(\chi\cdot A).
\end{equation}
Invariance of the electromagnetic vector field may be taken to be that $\lie^{'}_{\chi}A=0$.
This is modified form of invariance requirement where one insists that the coordinate invariance be extended to 
allow for non- invariance that can be compensated by the gauge transformation of that vector field, that is 
the invariant gauge potential is vanishing upto a total derivative, $\lie_{\chi}A =\,d(\lambda_{\chi})$. The modified derivative $\lie^{'}_{\chi}$ is sometimes called
the  Lie- Maxwell derivative. 

Similar to the the case for electromagnetic fields, one may also demand the existence of a derivative for the Lorentz  
transformations. Note that such kind of derivatives are useful since in many cases, even if the spacetime 
has a Killing vector ($\chi^{a}$), $\lie_{\chi}g_{ab}=0$, but the co- tetrad is not lie dragged,
$\lie_{\chi}e^{I}_{a}\neq 0$, which happens since the tetrad also suffers Lorentz transformations in 
the process of being Lie dragged. To take care of this additional Lorentz transformations that the tetrad frame undergoes, one usually defines a Lie- Lorentz derivative, which is a derivative $\lie^{'}_{\chi}e^{I}_{a}=0$. If the spacetime contains a Killing vector, one can show that such a choice of derivative is always possible \cite{Jacobson:2015uqa,Fatibene}. Then, 
for these spacetime, in these tetrad frames, the action of a diffeomorphisms can be realised as local Lorentz transformations. Particularly, consider the case that $\chi^{a}$ is a Killing 
vector field and that it's action on $e^{I}$ is given by:
\begin{equation}\label{LLie}
\lie_{\chi}\, e^{I}=\epsilon_{(\chi)}{}^{I}{}_{J}\, e^{J},
\end{equation}
where $\epsilon_{(\chi)}{}^{I}{}_{J}$ is an infinitesimal Lorentz 
transformation (associated to $\chi^{a}$) and hence is antisymmetric in the Lorentz indices. It is 
simple to show that $\lie_{\chi}\, g_{ab}=0$ \cite{Fatibene,Jacobson:2015uqa}.

We determine the $\epsilon_{IJ}$ for the near horizon spacetime.  First, note that the near horizon metric is expanded in the following form for a Kerr-like isolated horizon:
\begin{eqnarray}\label{metric_expansion}
ds^{2}&=&-r^2d\tau^2+\frac{d r^2}{\kappa^2}
+2r^{2}N(\theta)d\tau d\phi+2r^{2}\bar{N}(\theta)d\tau d\theta \nn
&+&2r M(\theta)dr d\theta+2r\bar{M}(\theta)dr d\phi
+X(\theta)d\theta^2\nn
&+&2Y(\theta)d\theta d\phi+Z(\theta)d\phi^2+\mbox{higher~orders}
\end{eqnarray}
where $N(\theta),\bar{N}(\theta), \dots $ are functions of spacetime \cite{Dreyer:2013noa}. 

This metric explicitly shows the fall- off conditions on the fields. As may be noted, $\partial_\tau^a$ is the Killing vector field and 
when written in the inertial coordinates, directly translates to the boost vector field. The tetrads 
corresponding to these metrics may be chosen to be
\beq
e^0=rd\tau~+\mathcal O(r^2),~e^1=\frac{dr}{\kappa}+\mathcal O(r)
\eeq
and the velocity four vector is $\chi^a=(1/r)\partial_\tau^a$. To progress further, we need to obtain the connection coefficients. Using the metric (\ref{metric_expansion}), it is possible to find that the connections $\Gamma$ behave in the following fashion:
\begin{gather}
\Gamma^{t}_{rt}=\frac{1}{r},~~\Gamma^{t}_{tt}\sim o(r^2),~~\Gamma^{t}_{t\theta}\sim o(r^2),~~\Gamma^{r}_{rt}=0,\nn~~\Gamma^{r}_{tt}\stackrel{\Delta}{=}r\kappa^2,~~\Gamma^{t}_{\theta t}\sim o(r^3),~~\Gamma^{\theta}_{\theta t}\sim o(r^2),~~\Gamma^{\phi}_{\theta t}\sim o(r^2)\nn
\Gamma^{\theta}_{rt}=0,~~\Gamma^{\theta}_{tt}\sim o(r^2)
\end{gather}
It is simple to check that the tetrad fields suffer change when carried along by the observer. More precisely,
the tetrads transform and mix among themselves
\beq \label{old_tetrad_basis}
\chi^a\nabla_ae^b_0&=&=\frac{\kappa}{r}e^b_1+\mathcal O(r)e^b_0+\mathcal O(r)\\
\chi^a\nabla_ae^b_1&=&=\frac{\kappa}{r}e^b_0+\mathcal O(r)
\eeq
These tetrads are not parallely propagated by the observer. Such a result is not surprising since tetrad frames rotate when they are carried along by the observer. The ones which are parallely transported are constructed by a linear combination of these tetrads and may be given by:
\beq\label{tetrad_new_basis}
\acute{e}^{0}&=&e^0\,\cosh{\sigma\tau}+e^1\sinh{\sigma\tau}\\ \nonumber
\acute{e}^{1}&=&e^1\cosh{\sigma\tau} + e^0\sinh{\sigma\tau},
\eeq
where the $\sigma=(1/l_{0})$ is the boost parameter. These tetrads in eqn. (\ref{tetrad_new_basis}) are the ones which have a direct one- to- one
relationship with the tetrads on the horizon. More precisely, the observer (fixed outside at a distance $l_{0}$) can construct a tetrad basis $(\acute{e}^{0}_{a}, \acute{e}^{1}_{a})$ which is parallely propagated along 
$\chi^a$.
It is simple to observe that the action
of the Lie- Lorentz derivative on this transformed primed tetrad is given by
\begin{equation}\label{tetrad_transfor}
\lie_\tau e'^0=(1/l_{0})\,~e'^1,~\mbox{and}~\lie_\tau e'^1=(1/l_{0})\,~e'^0,
\end{equation}
which implies that, in the notation of (\ref{LLie}) and using the fact that $\partial_\tau^a$ 
is the boost Killing vector field, $\epsilon^{1}{}_{0}=1=\epsilon^{0}{}_{1}$ with the boost
parameter $\sigma=(1/l_{0})$.  Note that the The tetrad frames are boosted by diffeomorphisms just like a local Lorentz transformation in that precise sense (compare eqns. \eqref{LLie} and \eqref{tetrad_transfor}). Then, we get from the equation \eqref{Ham_charge} (where, now $\sigma$ plays
the role of the boost parameter instead of $\eta$),
\beq\label{Ham_charge_2}
\delta Q=\delta\left(\frac{ A}{8\pi G\,l_{0}}\right).
\eeq
which is the Hamiltonian charge corresponding to the boost transformation, as obtained by an observer at a distance $l_{0}$ from the horizon. Clearly, this value $A/8\pi G\,l_{0}$ is also the quasilocal energy as perceived by the observer. 
Hence we get that $E=A/8\pi G\,l_{0}$ which is exactly the result of \cite{Frodden:2011eb}. 

\section{U(1) gauge transformations of the Electromagnetic field}

For the electromagnetic fields, the symplectic structure has two terms, one from the 
bulk and one from the 
surface 
\begin{equation}
\Omega_{EM}(\delta_{1}, \delta_{2})= -\frac{1}{2\pi}\int_{M} \delta_{[1}A\wedge \delta_{2]}{}^{*}F  -\frac{1}{2\pi}\int_{{\Delta}} \delta_{[1}\Psi\,\delta_{2]}{}^{*}F
\end{equation}
Here, the scalar $\Psi$ 
is defined as $\lie_{\chi}\Psi=(\chi \cdot A)$. 
There is a caveat though: if the condition $\lie_{\chi}\,A=d(\lambda_{\chi})$ holds not only on the horizon but also on the entire bulk, then, the contribution from the electromagnetic field does not arise. Consider a gauge transformation $A\rightarrow A+d\varphi$. The contribution from the bulk term comes out to be 
\beq
\Omega(\delta,\delta_\varphi)=\frac{1}{4\pi}\int_{M}d(\varphi~\delta~^{*}F)=\frac{\varphi}{4\pi}\delta\int_{S_{\Delta}}~^{*}F
\eeq
where $\varphi$ is a constant on $\Delta$. The boundary contribution can be shown to be zero by the following argument.
\begin{gather}
d\Psi=(l.A)n+\alpha m+\bar\alpha\bar m\\
d\delta_\eta\Psi=0
\end{gather}
Therefore one can set $\delta_\eta \Psi=0$ on the horizon cross-sections. 

The boosts along the observers trajectory at a fixed distance ($l_0$) outside the horizon may however induce a gauge transformations of the electromagnetic field for a charged black hole. It therefore remains to be shown  that the contribution from the electromagnetic part of the action to the first law vanishes. The electromagnetic field in the neighbourhood of the horizon is $A_a=\Phi (d\tau)_a$, $\Phi$ is a constant (using the metric \ref{metric_expansion}). It follows that to the leading order,
$ \chi^a\nabla_aA_b=\frac{\Phi}{r^2}(dr)_b$. By similar arguments as in the previous section, 
one should choose a $A'_a=A_{a}+\partial_a\varphi$, such that $\chi^a\nabla_aA'_b=0$. The choice of
$\varphi$ for this condition to hold is $d\varphi=\Phi dt$. It therefore follows that $\lie_\tau A'=\lie_\tau A=0$.

\section{Discussion}
In this paper, we have shown that (a) the energy of a black hole as determined by an observer at a distance $l_{0}$
from the horizon is $A/8\pi G l_{0}$, and (b) that the first law for these observers is given by  
$\delta H=\kappa\delta A/8\pi G $, with $\kappa$ being an universal constant for these observers 
with value $1/l_{0}$.  These are actually based on the result (which we derive in section 3) that the 
horizon area is the Hamiltonian charge for local Lorentz boosts and (as shown in section 4) that such Lorentz boosts 
may also arise out of the action of boost isometries on fields with internal Lorentz indices. These are 
the main results of our paper.

%

Some further comments are in order. Firstly, we emphasize that these results are obtained as boundary charges 
arising out of invariance of boundary conditions under Lorentz transformations.
It must also be pointed out that the quasilocal energy may also be derived from the boundary terms which 
survive on the horizon in the second order metric formulation. This may be seen as follows: the 
boundary Hamiltonian on a cross- section of a null surface due to an infinitesimal coordinate transformation  $x_{a}\rightarrow x_{a}+\epsilon_{a} (x)$ is given by:
%
$H=(1/ 8\pi G) \int d^{2}\Sigma^{ab}\, \nabla_{[a}\,\epsilon_{b]},$
%
where $d^{2}\Sigma^{ab}$ is an element of the cross- section and $\epsilon_{a}$ is the diffeomorphism
generated by the Killing vector $\chi^{a}$, which in the inertial frame of the Rindler 
spacetime is given by $\chi^{a}=(x/l_{0})\left[\left(\partial/\partial t\right)^{a}
+\left(\partial/\partial x\right)^{a}\right]$. In the inertial coordinates, the value of 
the surface element is ${1\over 2}\sqrt{q}\,d^{2}x_{\perp}$ and $\nabla_{[a}\,\epsilon_{b]}$ gives a value 
of $1/l_{0}$. Thus, the Hamiltonian is $A/8\pi G l_{0}$. Note that since this is a second order metric formulation,
it does not sense the action of local Lorentz transformations. On the other hand, our derivation of the local Lorentz boost and the horizon area
 is in the first order formalism. This gives an alternate perspective to the quasilocal first law. 
Thirdly, the limiting case of
extremal black holes is not easy to treat using the method described here since the Rindler description 
of the near horizon structure itself breaks down in that case. However since eqn. (\ref{boost charge}) is independent of whether the isolated horizon is extremal, one can argue that a first law holds for extremal isolated horizon with $\eta$ being an undetermined constant. Fourth, the result 
of \cite{Frodden:2011eb} assumes the existence of the first law in a certain form, which indeed arises 
if one considers general relativity (their subsequent construction, however, is independent 
of specific nature of the theory of gravity), but may be entirely different if one considers 
other theories of gravity, for example a scalar tensor theory. Then, it remains to see how
 the arguments of \cite{Frodden:2011eb} may change and what new results may arise. Furthermore, if 
 one considers the additional Holst term, one may also get contributions from the rotation part of 
 the Little group, and may have important implications for the simplicity constraints and the quantisation. These and
 other related matters will be discussed in a subsequent paper.


\subsection*{Acknowledgements}
The authors acknowledge the discussions with Amit Ghosh. 
AC is partially supported through the UGC- BSR start-up grant vide their letter no. F.20-1(30)/2013(BSR)/3082.
AG was supported by Department of Atomic-Energy, Govt. Of India. AG is currently supported by SERB, Government of India through the National Post Doctoral Fellowship grant (PDF/2017/000533).

\end{document}